\documentclass[10pt,twocolumn,a4paper]{article}

\usepackage[margin=2cm]{geometry} 
\usepackage{authblk}              
\usepackage{soul}                 
\usepackage[utf8]{inputenc}
\usepackage{hyperref}             
\usepackage{graphicx}             
\usepackage{verbatim}             

\usepackage{booktabs}
\usepackage{tabularx}

\usepackage[style=numeric, backend=biber, sorting=none]{biblatex}
\title{Human learning is an understudied but promising lever for boosting human--AI synergy}
\date{\today}                     
\author[a,b,c,d,1]{Julian Berger}
\author[e,f,a]{Jason W. Burton}
\author[a]{Ralph Hertwig}
\author[g]{Thomas Kosch}
\author[a]{Ralf H. J. M. Kurvers}
\author[h]{Benito Kurzenberger}
\author[g]{Christopher Lazik}
\author[h]{Linda Onnasch}
\author[h]{Tobias Rieger}
\author[a]{Anna I. Thoma}
\author[a,i,j]{Dirk U. Wulff}
\author[a,1]{Stefan M. Herzog}

\affil[a]{Center for Adaptive Rationality, Max Planck Institute for Human Development}
\affil[b]{Department of Business and Management, University of Southern Denmark}
\affil[c]{Department of Economics and Business, Universitat Pompeu Fabra}
\affil[d]{Barcelona School of Economics}
\affil[e]{Department of Psychology, University of Copenhagen}
\affil[f]{Copenhagen Center for Social Data Science, University of Copenhagen}
\affil[g]{Institut für Informatik, Humboldt-Universität zu Berlin}
\affil[h]{Department of Psychology \& Ergonomics, Technische Universität Berlin}
\affil[i]{Department of Psychology, University of Basel}
\affil[j]{Department of Business Analytics and Decision Sciences, Vienna University of Economics and Business}

\affil[1]{Correspondence: \texttt{berger@mpib-berlin.mpg.de} \& \texttt{herzog@mpib-berlin.mpg.de}}

\begin{document}

\maketitle

\begin{abstract}
Humans collaborating with artificial intelligence (AI) hold the promise of achieving superior outcomes compared to either acting alone (i.e., human--AI synergy). However, the conditions that facilitate such synergy when humans are advised by AI are not well understood.
A recent meta-analysis showed that, on average, human--AI combinations do not outperform the better individual agent. We argue that this pessimistic conclusion arises from insufficient attention to human learning in experimental designs. To substantiate this claim, we re-analyzed all 74 studies included in the original meta-analysis and found that most previous research overlooked design features that foster human learning (e.g., outcome feedback to participants). Our re-analysis further revealed that studies providing outcome feedback show tentatively higher synergy than those without outcome feedback. Crucially, feedback paired with AI explanations was associated with positive synergy, while explanations without feedback were associated with negative synergy---suggesting that explanations improve synergy mainly when humans can learn to verify the AI's reliability through feedback. Our re-analysis suggests that the current literature underestimates the potential of human--AI collaboration because it predominantly relies on paradigms that do not facilitate human learning, thus hindering humans from effectively adapting their collaboration strategies. However, experiments directly varying learning opportunities are needed for stronger, causal conclusions. We advocate for a paradigm shift in human--AI interaction research that explicitly addresses human learning and thus enhances our understanding of and support for successful human--AI collaboration.
\end{abstract}

\vspace{0.5em}
\vspace{1em}

\section*{Main}

\begin{figure*}[!t]
\centering
\includegraphics[width=\textwidth]{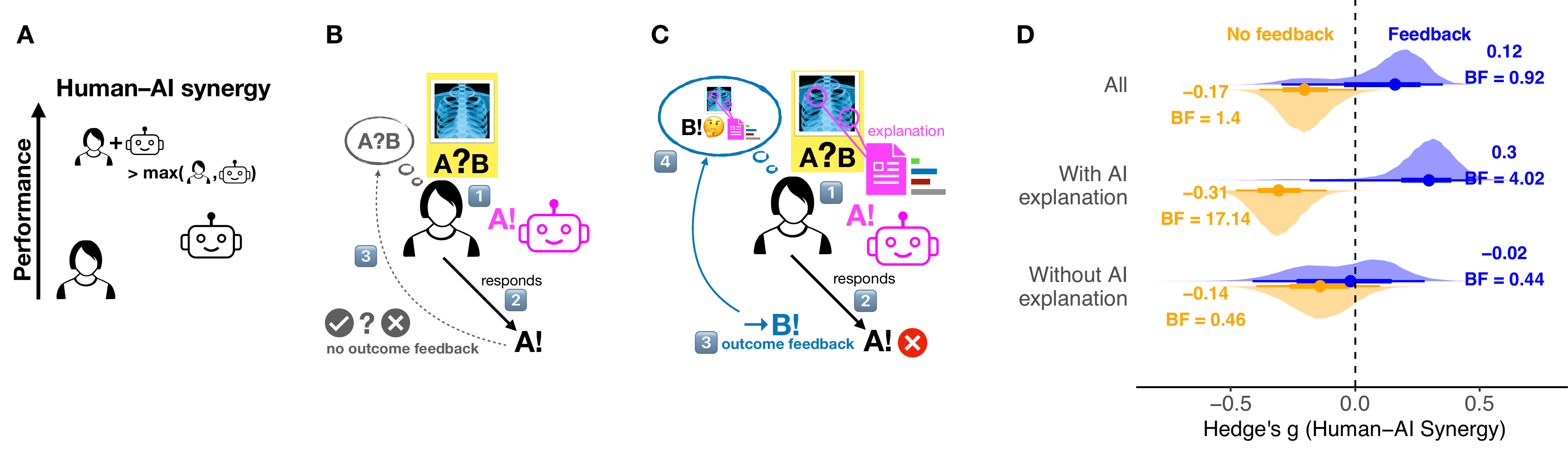}
\caption{ (A) Positive human--AI synergy occurs when the joint system outperforms either agent alone. (B) Human--AI interaction without learning opportunities: the human knows the AI's choice but receives no outcome feedback. (C) With learning opportunities: the human sees the AI's choice and its explanation, receives outcome feedback, and across trials learns about their own and the AI's relative skill. (D) Robust Bayesian model-averaged meta-regression analyses \cite{bartos_robust_2025} quantifying evidence that outcome feedback moderates human--AI synergy in the studies reviewed by Vaccaro et al. \cite{vaccaro_when_2024}. 
X-axis shows conditional marginal posterior estimates of Hedges' $g$ for studies with and without feedback. Dots and whiskers show the posterior median and 66$\%$/95$\%$ credible intervals. Bayes factors (BF) quantify evidence that the estimate is non-zero versus zero. Results are presented for all studies, and separately for studies with and without AI explanations.}\label{fig:results}
\end{figure*}

Humans collaborating with AI have the potential to achieve better outcomes than either independently, a phenomenon called human-AI synergy (Fig. \ref{fig:results}A). Synergy is well documented whenever independent human and AI judgments are combined statistically \cite{steyvers_bayesian_2022}. However, when and why this synergy occurs when humans interact with AI output is currently not well understood \cite{steyvers_three_2023}, although decision-theoretic analyses have characterized the bounds of possible synergy \cite{guo2024decision}.
Vaccaro et al. \cite{vaccaro_when_2024} recently conducted a systematic review and meta-analysis of 74 studies comparing the performance of humans, AI and human--AI combinations where humans needed to interact with AI systems. The studies span a wide range of tasks, such as classifying images, predicting house prices or writing code. On average, human--AI combinations performed worse than the best of humans or AI alone (i.e., negative human--AI synergy). Positive synergy emerged when humans outperformed AI alone, but synergy was negative when AI outperformed humans alone. Other moderators could not explain differences across studies. For instance, neither AI confidence nor AI explanations (helping humans understand how a model arrived at its output) led to human--AI synergy. We argue that human learning is a crucial, but overlooked facilitator of synergy (Figure \ref{fig:results}B and C). Re-analysing all studies covered in Vaccaro et al. \cite{vaccaro_when_2024}, we show that the few studies that gave humans feedback tended to report positive synergy---especially when AI explanations were also provided. We argue that neglecting human learning leads to underestimating the potential for human--AI synergy and therefore call for a greater focus on human learning to better understand and foster successful human--AI collaboration.

Conclusions about human performance and rationality drastically depend on how a phenomenon is studied \cite{almaatouq_beyond_2024,lejarraga_how_2021}. For example, when people can experience the probabilistic structure of a task environment through repeated interactions, their decisions are close to what probability theory would suggest whereas this is not so when participants are merely given textual descriptions of the task \cite{lejarraga_how_2021}. Could Vaccaro et al.'s \cite{vaccaro_when_2024} findings of negative synergy be partly due to the way human--AI collaboration is typically studied?

Learning, experience, and feedback are crucial for successful human decision making \cite{karelaia_determinants_2008,lejarraga_how_2021}. 
For example, observing outcome feedback (i.e., the correct response after each case) helps people identify reliable sources of high-quality information 
\cite{novaes_tump_individuals_2018}. 
It can also help humans assess the accuracy of their own decisions and of AI advice, and how best to integrate them into their final judgments \cite{jiangLearningHumanAICollaboration2025}. Also, guiding humans’ attention to fewer, relevant task features (e.g., a subset of medical diagnostic cues) and indicating how to use those features can further improve human judgments \cite{karelaia_determinants_2008}.
Explainable AI may likewise help humans focus on relevant task features---and even indicate how the AI uses those features---and thus facilitate human learning \cite{herzog_boosting_2024}. Such guidance should be particularly useful when humans and AI excel at different kinds of cases and humans need to learn to adaptively rely on their own judgment or AI's assistance \cite{rieger_explaining_2025}.
AI explanations, expressions of AI confidence and other meta information can thus serve as additional task features on top of humans’ own perceived confidence that can help humans develop adaptive strategies for when (not) to rely on AI \cite{steyvers_three_2023}.

To investigate how much human learning is recognized in the literature on synergy, we re-analyzed the studies reviewed by Vaccaro et al. \cite{vaccaro_when_2024}, quantifying design factors likely to benefit or hinder learning (e.g., provision of feedback or practice trials; coded separately for each experimental condition, see SI).
%
The typical study involved a single session with no practice trials, a median of 30 task trials and no outcome feedback---lack features known to facilitate learning.
Focusing on the 10 (14\%) of 74 studies that included experimental conditions with outcome feedback (providing 98 [26\%] of 370 effect sizes, see SI Extended Methods for details), only three studied how humans' use of AI assistance evolved across trials; the rest reported average performance across trials, potentially masking how humans adapt to the task based on feedback. None of the studies experimentally varied whether outcome feedback was provided or not.

To test whether outcome feedback---especially with AI explanations---fosters human--AI synergy, we ran robust Bayesian meta-regressions \cite{bartos_robust_2025} modeling synergy based on whether outcome feedback and AI explanations were provided (see SI). 
We 
quantify synergy using Hedges' $g$ effect size to measure the standardized difference in performance between the human--AI combination and whichever performed better alone. That is, positive (negative) estimates indicate that the human--AI combination performed better (worse) than either humans or AI alone. 
Comparing studies with (98 effect sizes) and without feedback (272; Fig. \ref{fig:results}D), provided tentative evidence that studies with feedback reported higher synergy ($BF_{inclusion}$ = 2.36).
Without feedback, synergy tended to be negative (Hedges' $g = -0.17$, 95$\%$ credible intervals [CI] $-0.37$ to 0.00, $BF_{inclusion}$ = 1.4) while positive with feedback ($g = 0.12$, 95$\%$ CI $-0.28$ to 0.34, $BF_{inclusion}$ = 0.92).
While neither estimate was credibly different from zero, there was tentative evidence that synergy was higher in studies with than without feedback (contrast: $g = 0.34$; 95$\%$ CI $-0.01$ to  0.69, posterior probability of a positive direction: 84$\%$).

Next, we investigated whether outcome feedback increased synergy when combined with AI explanations.
For studies with explanations (Fig. \ref{fig:results}D), feedback was a strong positive moderator of synergy ($BF_{inclusion}$ = 21). When people had access to both explanations and feedback (24 effect sizes, 14\%), synergy tended to be positive
($g$ = 0.3, 95$\%$ CI $-0.18$ to 0.48, $BF_{inclusion}$ = 4.02); when they had explanations but not feedback (148), synergy was clearly negative ($g = -0.31$, 95$\%$ CI $-0.48$ to $-0.11$, $BF_{inclusion}$ = 17.14).
With explanations, providing feedback was associated with increased synergy (contrast: Hedges' $g = 0.60$, 95$\%$ CI 0 to  0.94, PD = 97$\%$); without explanations, feedback did not moderate synergy ($BF_{inclusion}$ = 0.59) (Fig. \ref{fig:results}D). 
This result indicates that AI explanations increase synergy mainly when humans can learn to verify the AI's reliability.
To assess whether study characteristics confounded our analyses, we ran additional meta-regressions that included each moderator considered by Vaccaro et al. alongside feedback (see \href{https://osf.io/mqw4y/files/mgxzv}{https://osf.io/mqw4y}). 
The effect directions remained consistent across all 12 models.

In sum, our re-analysis of Vaccaro et al. \cite{vaccaro_when_2024} showed that the studies largely neglect human learning. This is consistent with Lai et al. \cite{lai_towards_2023} who surveyed 124 studies on human--AI decision making (after 2018 focused on human--computer interaction; HCI) and found that only 6 (5\%) of studies offered outcome feedback.
To better assess the role of human learning in human--AI collaboration through evidence synthesis, we make three recommendations:
First, we echo Vaccaro et al.'s \cite{vaccaro_when_2024} call to report standardized performance metrics, even when it is not the study's primary outcome.
Second, we call for sharing participant- and trial-level data, which would allow more directly studying how humans learn to use AI; we note a dearth of publicly available datasets.
Third, broader search terms and including studies prior to 2018 would better represent further relevant fields such as human factors \cite{rieger_explaining_2025} and judgment and decision making \cite{burton_systematic_2020}.
Three caveats:
First, the usefulness of feedback is constrained by the task structure and learning environment \cite{hogarth_two_2015, hullman_underspecified_2025}: When feedback is accurate, timely, and predictive task features are available, learning is facilitated; when feedback is noisy, delayed, or predictive task features are missing or opaque, learning is obstructed \cite{hogarth_two_2015}. As researchers design the task environment---for example, by (not) providing trial-by-trial feedback and AI explanations that guide attention to relevant features---the extent to which one observes human--AI synergy will depend on those choices.
Second, most studies only report average performance across trials, thus our between-study analyses cannot capture potential within-study improvements through feedback. Meta-analyzing joint performance in later trials---after participants have had a chance to learn---would better reflect the achieved learning.
Thus, our results may actually represent a conservative estimate of how learning may foster synergy.
We investigated such within-study learning in three studies where trial-level data was available (see \href{https://osf.io/mqw4y/files/uk3sj}{https://osf.io/mqw4y}) and found that participants learned to align with better-performing AI over trials when informative signals were available, but not otherwise.
Finally, our results are correlational and experiments directly varying learning opportunities are needed for stronger, causal conclusions.

Within the design space of possible human--AI decision making experiments, providing feedback or other learning opportunities is just one of many design choices. Task selection, timing and format of AI advice, incentives, participant populations, and evaluation metrics all jointly define the region of the design space from which data is collected and conclusions are drawn \cite{almaatouq_beyond_2024}. Hullman et al. \cite{hullman_underspecified_2025} argue that conclusions about human bias in AI advice taking are overly pessimistic and unwarranted because HCI studies---including some covered in Vaccaro et al. \cite{vaccaro_when_2024}---do not allow participants, or even idealized rational decision makers, to identify the correct response.
The pessimistic view of human--AI synergy \cite{vaccaro_when_2024} may thus reflect sampling from a learning-handicapping part of the design space rather than the frontier of possible human--AI synergy.
Going forward, we therefore advocate for a systematic and theory-informed exploration of the design space \cite{almaatouq_beyond_2024,hullman_underspecified_2025}, including experimentally varying the presence and type of feedback \cite{ahn_impact_2024} and the task structure \cite{karelaia_determinants_2008,hogarth_two_2015}. 


In sum,
our re-analysis suggests that supporting human learning---particularly by providing outcome feedback and AI explanations---is an underappreciated and understudied but promising lever for increasing human--AI synergy.
Considering the broader literature on human learning \cite{karelaia_determinants_2008,lejarraga_how_2021, novaes_tump_individuals_2018}, 
boosting competences \cite{herzog_boosting_2025,herzog_boosting_2024},
and the importance of study design \cite{almaatouq_beyond_2024,lejarraga_how_2021,hullman_underspecified_2025}, these meta-analytic results thus paint a more optimistic picture and highlight the need to explicitly and experimentally study when and how fostering human learning boosts human--AI synergy.

\subsection*{Data} 
We use open data from \cite{vaccaro_when_2024} for coding of effect sizes and AI explanations of 370 experimental conditions in the 74 studies analyzed. Other features (e.g., feedback) were coded by independent systematic review of every study by two of six authors (J.B., J.W.B., T.R., C.L., B.K., T.K.), with disagreements arbitrated by J.B. (see SI for coding scheme). 
\vspace{-6pt}
\subsection*{Meta-regression analyses}
All meta-regression models were estimated using RoBMA version 3.5. \cite{bartos_robust_2025}. We quantify human--AI synergy using Hedges' $g$ to measure the standardized difference in performance between the human--AI combination and whichever performed better alone. Positive (negative) estimates indicate that the human--AI combination performed better (worse) than either humans or AI alone. For more details on RoBMA models see SI.

\vspace{-9pt}
\subsection*{Data and code availability}
Data, code and fitted models are available at \href{https://osf.io/mqw4y/}{https://osf.io/mqw4y}.

\subsection*{Acknowledgments}
This work was supported by the European Union’s Horizon Europe research and innovation programme within the context of the project Hybrid Human Artificial Collective Intelligence in Open-Ended Domains (GA 101070588). This work was supported by the Deutsche Forschungsgemeinschaft (DFG, German Research Foundation) – grant HE 2768/11-1. The authors thank Susannah Goss for editing the manuscript. The authors thank František Bartoš for guidance and support on using RoBMA.

\section*{Author Contributions}
{\small
Conceptualization: J.B., T.R., J.W.B., A.I.T., L.O., R.H., R.H.K., and S.M.H. Data curation: J.B., T.R., J.W.B., C.L., B.K., T.K., and S.M.H. Formal analysis: J.B. and S.M.H. Investigation: J.B., T.R., J.W.B., C.L., B.K., A.I.T., D.U.W., L.O., R.H., R.H.K., T.K., and S.M.H. Methodology: J.B., T.R., J.W.B., C.L., B.K., T.K., and S.M.H. Project administration: J.B. and S.M.H. Resources: R.H., R.H.K., and S.M.H Software: J.B. Supervision: J.B. and S.M.H. Validation: J.B., T.R., and S.M.H. Visualization: J.B., D.U.W., and S.M.H. Writing - original draft: J.B. and S.M.H. Writing - review \& editing: J.B., T.R., J.W.B., C.L., B.K., A.I.T., D.U.W., L.O., R.H., R.H.K., T.K., and S.M.H.
}

\section*{Competing Interests}
{\small The authors do not declare any competing interests.}

\printbibliography

\clearpage

\onecolumn 

\setcounter{figure}{0}
\setcounter{table}{0}
\renewcommand{\thefigure}{S\arabic{figure}}

\section*{Supplementary Information}

\subsection*{Methods}

\subsubsection*{Data}

Vaccaro et al. \cite{vaccaro_when_2024} provided only human-only, AI-only and human--AI performance averaged across all trials of a study condition. While it would be ideal to test our research questions about human learning as a function of how many trials a participant has already engaged with, only three of the 10 studies with feedback reported results as a function of trial number. Thus, the data aggregated by Vaccaro et al. does not allow for such detailed individual-level analysis of potential learning effects. 

Moreover, only some studies in Vaccaro et al. openly share their data, and even fewer provide the data needed to reconstruct how participants chose across trials (e.g., \cite{poursabzi-sangdeh_manipulating_2021} Individual participant-level meta analysis remains our overall goal, but it requires a new systematic review that treats open data and clear trial sequences as inclusion criteria.
In the meantime, we present the within-study analyses of three published datasets in our OSF repository available at \href{https://osf.io/mqw4y/}{https://osf.io/mqw4y} and show that learning is both detectable and that accounting for it paints a less pessimistic picture of how humans and machines perform together.

\subsubsection*{Study design features}

Across the 370 effect sizes in \cite{vaccaro_when_2024}, we coded the following additional moderators during our systematic review. 
\begin{itemize}
    \item For the presence or absence of feedback during the main task (\texttt{feedback\_main}), 272 effect sizes had no feedback while 98 had feedback. 
    \item Among cases where feedback was provided (\texttt{feedback\_main\_n}), 94 effect sizes involved feedback on every trial and 4 involved feedback every 30 trials. 
    \item Practice trials were available (\texttt{practice\_available}) in 155 effect sizes and unavailable in 215. 
    \item The number of trials participants completed in the main tasks (\texttt{number\_responses}) had a median count of 30 (IQR = 15 to 100), and a mean count of 70.8 (SD = 80.12).
    \item All of the 370 effect sizes involved single-task scenarios as opposed to participants engaging in multiple tasks at once (\texttt{multitask}) and were conducted in artificial laboratory or online settings (\texttt{setting}) as opposed to natural or field experiments. 
    \item Regarding the number of experimental sessions (\texttt{sessions}), 332 effect sizes came from studies with one single session while 38 came from studies that invited participants into multiple sessions. 
        \item With regard to performance incentives (\texttt{incentive}), 88 effect sizes involved no incentive, 141 involved performance-based incentives, and 141 provided no information about incentives. 
    \item The level of automation (\texttt{automation}) varied: 283 effect sizes involved systems providing one specific solution, 85 involved systems providing multiple solutions, and 2 involved systems showing and implementing solutions unless vetoed by the human participant. 
    \item Finally, the sequence of human–AI decision making (\texttt{sequence}) was concurrent in 236 effect sizes (i.e., participants deciding while observing system output), sequential in 114 (i.e., participants deciding independently first before being able to revise with AI), on-demand in 8 (i.e., participants being able to request assistance), time-delayed in 8 (i.e., participants needing to wait for some time before seeing system output), and involved the system offering advice when deemed necessary as judged by the AI in 4 instances.
\end{itemize}

\subsubsection*{Studies with feedback}

Ten studies provided performance feedback in the main task, contributing 98 effect sizes across 14 different tasks. Most tasks involved classification or prediction decisions: \cite{alufaisan_does_2021} had participants predict recidivism with and without AI explanations (total number of effect sizes contributed to meta-analyses, n = 12); \cite{bansal_does_2021} examined classifying beer reviews, Amazon reviews, and answering LSAT questions with and without AI explanations (each n = 12); \cite{boskemper_measuring_2022} studied baggage x-ray screening without AI explanations (n = 3); \cite{elder_knowing_2024} examined basketball game predictions without explanations (n = 6); \cite{he_knowing_2023} studied image labeling with and without AI explanations (n = 5); \cite{kyriacou_learning_2021} examined sorting recyclables without explanations ( = 1); \cite{tejeda_ai-assisted_2022} studied image classification without explanations (n = 48); \cite{yang_how_2020} examined leaf and liver image classification with and without explanations (each n = 6); and \cite{zhang_you_2022} studied shape classifications without explanations (n = 8). Only one study involved a creation task: \cite{weisz_better_2022} examined code translation without AI explanations (n = 6).

\subsubsection*{Robust Bayesian Model Averaging}

We used a Robust Bayesian Model Averaging (RoBMA; Rpackage version 3.5) approach for our meta-regressions  evaluating moderation effects \cite{bartosRobustBayesianMultilevel2025, bartos_robust_2025}. This approach runs many models, systematically varying assumptions (e.g., absence versus presence of main or moderator effects, absence versus presence of publication bias) and then calculates a posterior effect-size estimate based on the average of all models, weighted by how well the models account for the data. 
The resulting meta-regression models quantify the evidence for moderation using Bayes factors ($BF_{inclusion}$) and make it possible to calculate the conditional marginal posterior effect-size estimate for different moderator levels (i.e., feedback versus no feedback). 

We used R 4.4.2. for all analyses. Our RoBMA models have the same three-level hierarchical model structure originally used by Vaccaro et al. \cite{vaccaro_when_2024}, nesting effect sizes within experiment and experiments within study. We relied on default priors as specified by the RoBMA authors \cite{bartos_robust_2025}. 
Detailed model summaries and fitted model objects are provided in the project repository available at \href{https://osf.io/mqw4y/}{https://osf.io/mqw4y}. See Bartoš et al. \cite{bartosRobustBayesianMultilevel2025,bartos_robust_2025} and \href{https://fbartos.github.io/RoBMA/}{https://fbartos.github.io/RoBMA/} for guidance on how to interpret those outputs.

We fit all models using the spike-and-slab algorithm. Models were run for at least 45,000 samples after 15,000 burn-in and 10,000 adaptive samples. In case models did not converge, burn-in and adaptive samples were increased up to 25,000 and 20,000 respectively. Convergence was checked using the Gelman--Rubin criterion $\hat{R}$ $<$ 1.05, relying on the automatic convergence checks implemented in RoBMA. 

To calculate the contrast between effect-size estimates of Hedge's $g$ for studies with versus without feedback, we extracted the conditional marginal posterior distributions from the RoBMA models and summarized the resulting posterior distribution of the difference in effect-size estimates using \texttt{tidybayes::median\_qi()} \cite{kay_tidybayes_2024}. We also describe the posterior difference using the probability of direction, using \texttt{bayestestR::p\_direction()} \cite{makowski_bayestestr_2019} to quantify how much of the posterior distribution of the difference is greater than 0.

\end{document}